\documentclass[conference]{IEEEtran}

\usepackage{cite}
\usepackage{amsmath,amssymb,amsfonts}
\usepackage{algorithmic}
\usepackage{algorithm}
\usepackage{graphicx}
\usepackage{textcomp}
\usepackage{xcolor}
\usepackage{url}
\usepackage{tikz}
\usepackage{caption}
\usepackage{subcaption}
\usepackage{balance}
\usepackage{float}

\usepackage{flushend}

\graphicspath{{images/}}

\begin{document}

\title{FastGFDs: Efficient Validation of Graph Functional Dependencies with Desbordante}

\date{}

\author{
\IEEEauthorblockN{Anton Chernikov\textsuperscript{1},  Yurii Litvinov\textsuperscript{1}, Kirill Smirnov\textsuperscript{1}, George Chernishev\textsuperscript{1,2},}
\IEEEauthorblockA{
\textsuperscript{1} Saint-Petersburg University \\ 
\textsuperscript{2} Universe Data \\
Saint-Petersburg, Russia\\ 
\textsuperscript{3}st068877@student.spbu.ru, \{y.litvinov, k.k.smirnov, g.chernyshev\}@spbu.ru\\
}
}

\maketitle

\begin{abstract}

Graph functional dependencies (GFD) are a recently-developed concept aimed at capturing both topological structures in graphs
and functional dependencies between attributes. The process of verifying whether a given GFD holds over a particular graph is referred to as GFD validation. In this very computationally expensive problem, locating suitable subgraphs accounts for about 99\% of the total run time. The concept's authors originally proposed a parallel scheme (algorithm), targeting specifically clusters of high-performance servers. 

The goal of this study is to open GFD validation to a broader public by making it possible to run it on a consumer class PC. Our initial experiments demonstrated that the existing algorithm may not be optimal for these purposes. Therefore, we propose FastGFDs~--- a GFD validation algorithm that employs a recently developed graph matching technique. In contrast to the parallel scheme, it is sequential and operates on the entire graph. Its novelty lies in the use of Core-First Decomposition and the Compact Path Index (CPI). We compare it with the naive sequential algorithm and the parallel scheme, evaluating run times and memory consumption. The current study is the first step towards designing an efficient algorithm for GFD validation in low-end single-node environments.

We also provide an open-source implementation of GFD validation over large data graphs. To the best of our knowledge, this is the only publicly available implementation of an algorithm for this problem. It is developed in Desbordante~--- an open-source high-performance data profiler aimed at science-intensive tasks.

Finally, our experiments on a real-life graph demonstrated up to three times performance (2.6x on average) improvement over the parallel scheme. Employing the new subgraph matching algorithm also reduced memory consumption by five times.

\end{abstract}

\section{Introduction}

Graph functional dependencies~\cite{fan2016functional} (GFDs) are a generalization of relational functional dependencies~\cite{10.1145/2882903.2915203} to graph data. They consist of a graph pattern and a set of functional dependencies between the labels of nodes and edges in the pattern. They have the following semantics: a graph functional dependency is satisfied if, for every occurrence of the pattern in the data graph, every functional dependency specified by the GFD holds. 

GFDs capture topological structures as well as functional dependencies between attributes and have many uses in graph databases, such as~\cite{fan2016functional, 10.1145/3397198}: searching for inconsistencies, defining schema for graph data, formulating graph integrity constraints, detecting spammers and fake accounts in social networks, and many more.

\begin{figure}
    \centering
    \includegraphics[width=0.4\textwidth]{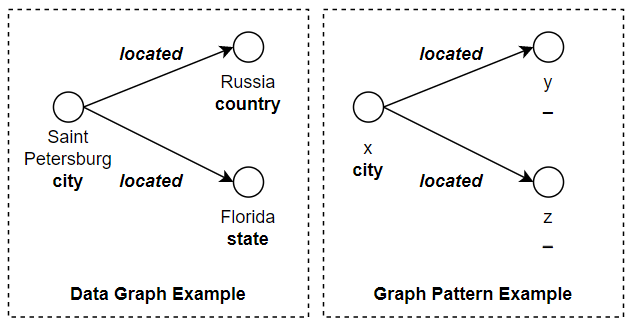}
    \caption{Data inconsistency in YAGO3 and the graph pattern that finds it.}
    \label{image:dataAndGfdExample}
\end{figure}

As an example, consider the inconsistency that Fan et al.~\cite{10.1145/3397198} found using GFDs in the YAGO3~\cite{mahdisoltani2014yago3} graph database and the graph functional dependency that was used to detect it, in Fig.~\ref{image:dataAndGfdExample}. Here, St. Petersburg is located in two places, Russia and Florida~--- in reality, they are two different cities, but they got incorrectly merged in YAGO3. This can be detected via a GFD with the graph pattern from the right side of Fig.~\ref{image:dataAndGfdExample} and an attribute dependency $\emptyset \rightarrow y.name = z.name$, where $\emptyset$ denotes an empty set of conditions in the ``if'' part and $y.name = z.name$ denotes that in a match of graph pattern from Fig.~\ref{image:dataAndGfdExample} names of locations must be the same (i.e. if a city is located in $y$ and $z$, then $y$ and $z$ are the same place). 

\begin{figure}
    \centering
    \includegraphics[width=0.25\textwidth]{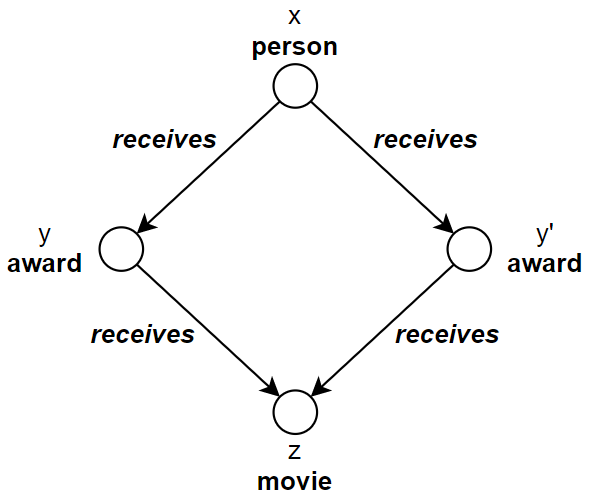}
    \caption{Graph pattern of a more complex GFD discovered in YAGO2.}
    \label{image:movieAwardGfd}
\end{figure}

Furthermore, GFDs can be automatically discovered in large graphs, thus helping to find out non-trivial facts about data. For example, GFD $Q[x, y, z, y'](y.name = ``Gold Bear" \wedge y'.name = ``Gold Lion" \rightarrow
false)$ was discovered~\cite{10.1145/3397198} in the YAGO2 database, and its graph pattern can be seen in Fig.~\ref{image:movieAwardGfd}. It means that no movie has ever received both the Golden Bear and the Golden Lion awards, which is correct, since both festivals require movies to be an initial release.

As seen in examples above, there are two core user-facing problems of GFDs:
\begin{enumerate}
    \item GFD validation~\cite{fan2016functional}~--- for a given GFD $\varphi$ and a graph $G$, check that dependencies in $\varphi$ are not violated in $G$.
    \item GFD discovery~\cite{10.1145/3397198}~--- for a given graph $G$, find all dependencies that hold in $G$.
\end{enumerate}

This study is focused on the GFD validation problem. It is very computationally expensive: it was shown that it belongs to the co-NP-complete class~\cite{fan2016functional}. However, there are scalable parallel algorithms that make it possible to check GFDs in real-world large graphs in reasonable time~\cite{fan2016functional}. Unfortunately, the original paper provided no implementation and to the best of our knowledge, there is no publicly-available implementation of this algorithm. 

The central issue of the GFD validation is the graph pattern matching complexity. Our preliminary experiments demonstrated that locating suitable subgraphs takes up about 99\% of the total run time and only 1\% is spent on checking the dependency. However, recent advances in algorithms and data structures for graph pattern matching, such as~\cite{bi2016efficient} and~\cite{yang2011br}, make it possible to speed up this part. 

In this paper, we present FastGFDs~--- a fast algorithm for validation of GFDs. Its novelty is the use of Core-First Decomposition and the Compact Path Index (CPI) described in paper~\cite{bi2016efficient} intended to speed up the subgraph matching part of the original algorithm. 

This technique aims to reduce the number of redundant graph Cartesian products. The idea is to decompose the query graph into a dense subgraph and a forest, and then process the dense subgraph first. This dense subgraph will have more edge-connectivity information, which can be exploited during matching. The matching itself is performed using the CPI, which allows to avoid enumerating all embeddings of a query path in the data graph.

The original paper~\cite{fan2016functional} did not provide explicit details about the subgraph matching algorithm. Instead, one of the core contributions was the introduction of a parallel scheme that included subgraph matching. Roughly speaking, its idea is assigning work units to multiple processors in a balanced manner. Each work unit is responsible for matching a pattern in a graph chunk. In contrast to this, FastGFDs uses a sequential matching algorithm that operates on an entire graph.

The authors of the parallel scheme benchmarked it using a cluster of high-end servers with nearly two hundreds of processors. Their experiments demonstrated good scalability and in general proved its suitability for massively parallel environments. Our idea is to study the performance of the original parallel scheme in a local multi-threaded case with low-end hardware, evaluating both its performance and memory usage. We aim to assess the necessity of the parallel scheme in such an environment and compare it to a modern sequential algorithm. This examination serves as the first step towards designing an efficient algorithm for GFD validation in low-end single-node environments.

Overall, we have implemented the original parallel algorithm, a naive sequential algorithm, and FastGFDs within the Desbordante platform~\cite{Desbordante}~--- an open source science-intensive data profiling tool. To the best of our knowledge, this is the only publicly available implementation of a GFD validation algorithm. We then conducted a series of experiments to study these algorithms.

In summary, our main contributions can be outlined as follows:

\begin{enumerate}
    \item we provide an open-source implementation of the baseline GFD validation algorithm from~\cite{fan2016functional} in C++ for maximum performance;
    \item we improve the original algorithm from~\cite{fan2016functional} using the ideas from~\cite{bi2016efficient};
    \item we present an experimental evaluation for our implementations of three algorithms.
\end{enumerate}

The rest of this paper is organized as follows. Section~\ref{sec:background} introduces main definitions and the used notation, and formalizes the problem. Section~\ref{sec:relwork} provides a brief overview of related work. Section~\ref{sec:solution} describes the proposed solution, Section~\ref{sec:evaluation} provides its evaluation on both a real-life graph. In Section~\ref{sec:threats} we discuss threats to the validity of our study. Finally, Section~\ref{sec:conclusion} concludes the paper.

\section{Background}
\label{sec:background}

In this section, we will formally define the concepts and notation, closely following~\cite{fan2016functional}.

We consider directed labeled graphs with attributes (we will refer to them as \emph{graphs} for brevity) and define them as follows: 
$G = (V, E, L, F_A)$, 
where $V$ is a finite set of vertexes, 
$E \subseteq V \times V$ is a set of edges, 
$L$ is a label function (i.e. $L(v)$ or $L(e)$ provides a label for a vertex $v \in V$ or an edge $e \in E$), 
$F_A$ is an attribute function that maps every vertex $v \in V$ to a tuple $(A_1 = a_1, \ldots, A_n = a_n)$ where $a_i$ is some constant. 
$A_i$ is an \emph{attribute} of vertex $v$, and we denote its value (i.e. $a_i$) as $v.A_i$. Note that edges do not have attributes.

We also define a \emph{subgraph} $G' = (V', E', L', F'_A)$, denoted by $G' \subseteq G$ as a graph that contains a subset of edges and vertices from $G$: $V' \subseteq V$, $E' \subseteq E$, $\forall v \in V\ L'(v) = L(v)$ and $F'_A(v) = F_A(v)$, $\forall e \in E\ L'(e) = L(e)$.

\begin{figure}[H]
    \centering
    \includegraphics[width=0.45\textwidth]{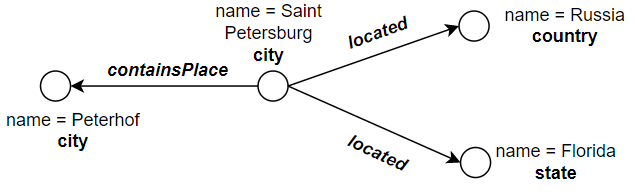}
    \caption{Directed labeled graph example.}
    \label{image:graphExample}
\end{figure}

An example of a directed labeled graph is shown in Fig.~\ref{image:graphExample}. Here, ``city'', ``country'' and ``state'' are vertex labels, ``located'', ``containsPlace'' are edge labels, ``name'' is an attribute, and ``Peterhof'', ``Saint Petersburg'', ``Russia'' and ``Florida'' are attribute values. The left part of Fig.~\ref{image:dataAndGfdExample} is a subgraph of this graph.

A \emph{graph pattern} is a graph $Q[\bar{x}] = (V_Q, E_Q, L_Q, \mu)$ where $V_Q$ is a set of nodes in a graph pattern, 
$E_Q \subseteq V_Q \times V_Q$ is a set of edges, 
$L_Q$ is a label function (with the same meaning as above), 
$\bar{x}$ is a list of variables for each vertex (i.e. $|\bar{x}| = |V_Q|$, and $\mu$ is a bijective map from $\bar{x}$ to $V_Q$, i.e. $\forall x \in \bar{x}\ \mu(x) = v \in V_Q$. 
Intuitively, $\bar{x}$ is a set of vertexes to be found in a data graph $G$ by pattern (or query) $Q$, and we will define graph pattern matching as filling variables from $\bar{x}$ with vertexes from $G$ so that they ``correspond'' to the pattern $Q$. We will use simply $x$ to denote $\mu(x)$ when it will not lead to confusion.

More precisely, a \emph{match} of a pattern $Q[\bar{x}]$ in a graph $G$ is a subgraph $G' = (V', E', L', F'_A)$ that is isomorphic to $Q[\bar{x}]$, i.e. there exists a bijective function $h: V_Q \rightarrow V'$ such that $\forall v \in V_Q\ L_Q(v) = L'(h(v))$ (i.e. labels match) and $e = (v, v') \in E_Q \iff e' = (h(v), h(v')) \in E'$ and $L_Q(e) = L'(e')$ (i.e. edges for matching vertices match --- note that a matching subgraph cannot contain ``additional'' edges). We also use the wildcard symbol $\_$ that matches any label, i.e. if $L_Q(v) = \_$ then $L_Q(v) = L'(h(v))$ always holds. And we will use $h(\bar{x})$ to denote a match of $Q[\bar{x}]$ in a graph $G$.

\begin{figure}
    \centering
    \makebox[0.45\textwidth][c]{\includegraphics[width=0.50\textwidth]{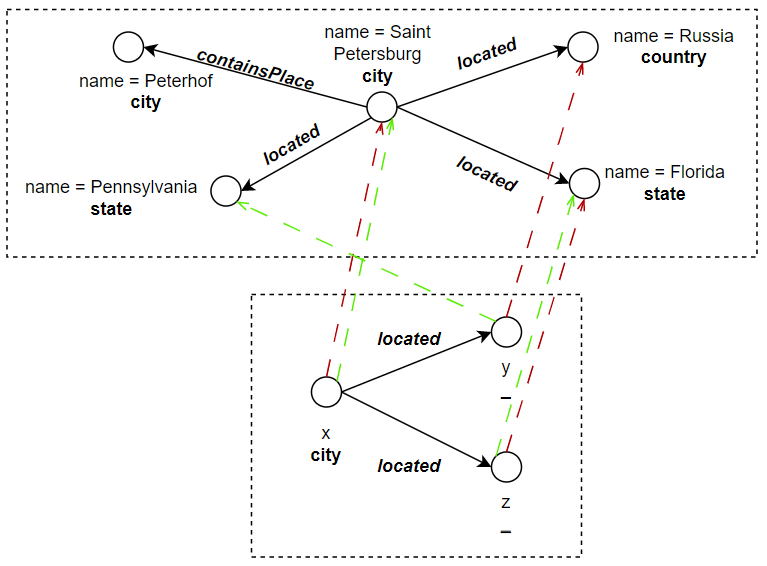}}
    \caption{Pattern and data graph match example.}
    \label{image:matchExample}
\end{figure}

An example of two of six possible matches is shown in Fig.~\ref{image:matchExample}. Here $G'_1$, a match denoted by red dashed arrows in Fig.~\ref{image:matchExample}, consists of vertices with names ``Saint Petersburg'', ``Russia'' and ``Florida'', edges incident to  them and corresponding labels and attributes. $\bar{x} = (x, y, z)$, $h(x)$ is a vertex with name ``Saint Petersburg'', $h(y)$~--- ``Russia'' and $h(z)$~--- ``Florida''.

Now we can formally introduce \emph{graph functional dependencies}. A GFD $\varphi$ is either a pair $(Q[\bar{x}], X \rightarrow Y)$ or a pair $(Q[\bar{x}], X \rightarrow false)$, where $Q[\bar{x}]$ is a graph pattern and $X$ and $Y$ are two sets of \emph{attribute constraints}. An attribute constraint is either a \emph{constant constraint} $x_i.A = c$ or a \emph{variable constraint} $x_i.A = x_j.B$ where $x_i$, $x_j \in \bar{x}$. An \emph{attribute dependency} $X \rightarrow Y$ is similar to a relational functional dependency and intuitively has a meaning of ``if for a match of graph pattern $Q[\bar{x}]$ all constraints from $X$ hold, then all constraints from $Y$ also hold''. More formally, given a match $h(\bar{x})$ of $Q[\bar{x}]$ in $G$, we say that $h(\bar{x})$ \emph{satisfies} a constant constraint $x_i.A = c$ if in a node $v = h(x_i) \in G$ there exists an attribute $A$ and $v.A = c$. Similarly, for variable constraints $x_i.A = x_j.B$: $h(x_i).A = h(x_j).B$. No match $h(\bar{x})$ can satisfy $false$. We denote $h(\bar{x}) \models X$ if $h(\bar{x})$ satisfies all the constraints in $X$.

For example, Fig.~\ref{image:matchExample} illustrates the GFD $\varphi = (Q[x, y, z], \emptyset \rightarrow \{y.name = z.name\}$. It has no ``if'' part, but it requires that for a match of $y$ and $z$ pattern nodes, their ``name'' attributes must be equal.

Graph $G$ \emph{satisfies} GFD $\varphi$ (denoted by $G \models \varphi$) if for all matches $h(\bar{x})$ of $Q$ in $G$ if $h(\bar{x}) \models X$ then $h(\bar{x}) \models Y$. 

In our example from Fig.~\ref{image:matchExample}, the graph $G$ satisfies GFD $\varphi$ if for any city that is connected with ``located'' edges with at least two vertices, those vertices will have the same value of the ``name'' attribute. If we recall that the YAGO database is derived from Wikipedia, WordNet and so on, it is possible that some relations come from different sources and are duplicated, but different targets for some relations point to errors in data.

Intuitively, a graph pattern provides a topological context for dependency, and for \emph{all} occurrences of this context in data graph functional dependencies over attributes must hold. Also, graph pattern defines an equivalent of a relational schema for GFDs, allowing to match attributes. But note that match in the data graph may not contain all attributes from $X$, in that case relevant constraints are holding trivially. This property of GFDs makes them useful for semi-structured data, which is common in graph databases. In contrast, by definition of satisfaction, if a constraint is in $Y$, it enforces the existence of all relevant attributes in a data graph, which allows to, for example, express typing constraints $(Q[x], \emptyset \rightarrow x.A = x.A)$ which require a vertex with some label defined in a single-vertex pattern $Q[x]$ have an attribute $A$.

At last, we can define the validation problem. Given a GFD $\varphi = (Q[\bar{x}], X \rightarrow Y)$ and a graph $G$, we say that a match $h(\bar{x})$ of $Q[\bar{x}]$ in $G$ is a \emph{violation} of $\varphi$ if $G_h \not\models \varphi$, where $G_h$ is a subgraph induced by $h(\bar{x})$ (i.e. there is a structural match, but FDs over attributes do not hold).

For a set $\Sigma$ of GFDs and a given graph $G$ we say that $G$ \emph{satisfies} $\Sigma$ (and denote it as $G \models \Sigma$ if there are no violations of GFDs from $\Sigma$ in $G$. \emph{The validation problem} for a set of GFDs $\Sigma$ and a graph $G$ is formulated as finding all violations of all GFDs from $\Sigma$ in $G$, i.e. given a set of GFDs $\Sigma$ and a graph $G$ find a set $Vio(\Sigma, G)$ of violations, where $Vio(\Sigma, G) = \{h(\bar{x})\ |\ \exists \varphi \in \Sigma: G_h \not\models \varphi \}$. 

In our example from Fig.~\ref{image:matchExample} $\varphi = (Q[x, y, z], \emptyset \rightarrow \{y.name = z.name\} \in Vio(\{\varphi\}, G)$, because there exist a match where $y.name = Russia$ and $z.name = Florida$ and $Russia \not= Florida$.

It is shown in reference~\cite{fan2016functional} that the validation problem is coNP-complete.

\section{Related Work}
\label{sec:relwork}

The current paper is based on~\cite{fan2016functional}, which defines and examines graph functional dependencies, states problems of satisfiability, implication and validation, and describes and evaluates validation algorithms on large real-life graphs. It is concluded that despite the coNP-hardness of the validation problem, it is possible to devise parallel-scalable algorithms (an algorithm is parallel-scalable if it works faster after adding more computation nodes), that have adequate performance and can actually be used for real-life graph analysis, and descriptions of two such algorithms are provided. However, paper~\cite{fan2016functional} does not provide an implementation of these algorithms, and our first difference from this work is actually providing a reusable implementation of one of the parallel GFD validation algorithms. Note that our work is much more technical than~\cite{fan2016functional} and has a limited scope~--- we do not consider satisfiability and implication problems at all, and we do not implement the graph partitioning algorithm from~\cite{fan2016functional}, but we extend and improve the ``replicated graphs'' algorithm, which can be considered as a baseline for our research.

There is a more recent work from the authors of~\cite{fan2016functional}, concerning the discovery of GFDs,~\cite{10.1145/3397198}. It shows that there is a parallel-scalable algorithm for GFD discovery that is feasible for real-life large graphs. However, we do not cover discovery problem in our study.

Since the most time-consuming part of GFD validation is finding a match for a graph pattern $Q[x]$ in a data graph $G$, it could be worthwhile to look at recent graph pattern matching studies:

\begin{itemize}
    \item Reference~\cite{bi2016efficient} provides a good overview of existing graph pattern matching algorithms and proposes to search for a query graph $Q$ in a data graph $G$ using heuristics that minimize required vertex comparisons. According to~\cite{bi2016efficient}, all graph pattern matching algorithms are heuristic due to the NP-hardness of the problem, and all follow Ullmann’s backtracking approach: find a vertex in a data graph with a label that matches some vertex of a query graph and then iteratively grow the partially matched subgraph following the edges from the query graph and finding matching edges in the data graph. The selection of vertices to consider and the order in which matching is done is extremely important~--- an algorithm needs to consider all possible matches for edges incident to a given vertex, which can lead to unnecessary matches when it is possible to decide that there is no match by considering vertices in another order.
    
    The main idea of~\cite{bi2016efficient} is to start matching from the ``core'' of a query graph, i.e. a set of highly connected vertices, postponing matching of vertices with fewer connections and especially leaf vertices (vertices with degree of 1). Highly-connected subgraphs with a given structure are ``rarer'' in data graphs, so this selection of matching order filters out negative matches earlier. The authors also propose a ``compact path-index'' (CPI) auxiliary data structure, which maintains a candidate set of vertices in $G$ for each vertex from $Q$ and a set of edges that match corresponding edges for candidate vertices from $G$. It is used to generate the matching order and to conduct matching that significantly reduces the amount of ``unpromising'' partial matches.

    Performance evaluation shows~\cite{bi2016efficient} that their technique outperforms existing graph pattern matching algorithms by up to three orders of magnitude on a real and synthetic graphs, so it is natural to try to apply it to the GFD validation problem. Note that it can not be applied directly, since in~\cite{bi2016efficient} additional edges are allowed in $G$ that are not present in $Q$, but both of their incident vertices are in $Q$. However, our definition of a graph pattern match explicitly prohibits it.
    
    \item The study~\cite{yang2011br} describes an updatable indexing structure for fast graph pattern matching. At first, authors build an index of small subgraphs (called \emph{features}) in a data graph and partition the data graph into overlapping regions that contain those features. Index maps features to regions they are contained in, and every vertex of the data graph into its regions. It is shown that using some storage optimizations such index can be relatively small for real-life large graphs and can be constructed in a reasonable amount of time. For example, it requires 2.2GB to store the Flickr website user accounts graph, with more than 1.2M vertices. Its built time is a bit less than two hours on a two 3.33GHZ quad-core CPUs and 32 GB of RAM.

    Searching for a pattern using that index begins with identifying features in a query graph (note that the query graph can contain several features), then finding these features in the index and getting candidate subgraphs by joining features, then checking each candidate as usual, by growing the matched graph.
\end{itemize}

\section{Proposed Solution}
\label{sec:solution}

We describe both the baseline algorithm, which is our implementation of the ``replicated graphs'' algorithm from~\cite{fan2016functional} and our improvement, FastGFDs. Both algorithms take the data graph $G$ and a set of GFDs $\Sigma$ as input, and produce the set of violations $Vio(\Sigma, G)$ as defined in Section~\ref{sec:background}.

\subsection{Baseline algorithm implementation}~\label{sec:baseline-algo}

Following~\cite{fan2016functional}, we define \emph{workload} $W(\Sigma, G)$ as a necessary amount of work required to compute $Vio(\Sigma, G)$. For an algorithm to be parallel scalable, we partition $W(\Sigma, G)$ so that we can assign an approximately equal amount of work to $n$ processors.

The algorithm itself is described in detail in study~\cite{fan2016functional}, and here we will provide a brief overview.

\begin{enumerate}
    \item Given coordinator $S_c$ and $n$ processors $S_1, \ldots, S_n$, we estimate the total workload $W(\Sigma, G)$ and create a balanced partition $W_i(\Sigma, G)$ for $i \in [1, n]$ in parallel.
    \begin{enumerate}
        \item The coordinator $S_c$ for each GFD $\varphi \in \Sigma$ creates a pivot vector $PV(\varphi)$ --- given a graph pattern $Q$ of $\varphi$, a pivot vector is a set of centers of connectivity components of $Q$, where centers are vertices with the minimum of the maximal distances between other vertices in the connectivity component. For example, for pattern $Q$ from Fig.~\ref{image:matchExample} ``Saint Petersburg'' is the center of its only connectivity component.
        \item Then it calculates the frequency distribution of \emph{candidate} vertices $C(z_i)$ in $G$ --- nodes in the data graph that have the same label as centers $z_i$ for the pattern $Q$.
        \item Next, it evenly partitions candidates $C(z_i)$ into $m$ sets so that the number of candidates with attribute values that fall into a range of each partition is even.
        \item Then it constructs a set of $m$ messages with $\varphi$ and partitions from the previous step and evenly distributes them to $n$ processors.
        \item Then each processor $S_i$ for each received message identifies work units by finding all candidate vertices with attributes falling in a given range and reports back to coordinator $S_c$ such candidates and the count of vertices in $G$ within radius equal to radius (maximum distance from the center) of the corresponding connectivity component.
        \item The coordinator uses this information to partition $W(\Sigma, G)$ to n pair-wisely disjoint work unit sets $\{W_1(\Sigma, G), \ldots, W_n(\Sigma, G)\}$ consisting of work units from the previous step, using a greedy algorithm to evenly distribute workload.
    \end{enumerate}
    \item Then each processor $S_i$ computes a local set of violations by finding matches $h(\bar{x})$ of pattern $Q$ and checking whether $h(\bar{x}) \models X \Rightarrow h(\bar{x}) \models Y$. All detected violations are sent back to $S_c$.
    \begin{itemize}
        \item Note that~\cite{fan2016functional} does not specify a matching algorithm for matching pattern $Q$ in $G$.
    \end{itemize}
    \item Then the coordinator $S_c$ merges all detected violations into $Vio(\Sigma, G)$ which becomes the output of the algorithm.
\end{enumerate}

Our implementation uses separate threads as processors, so there is very little communication overhead. 

We use the Boost Graph Library~\cite{siek2001boost} to represent graphs. 

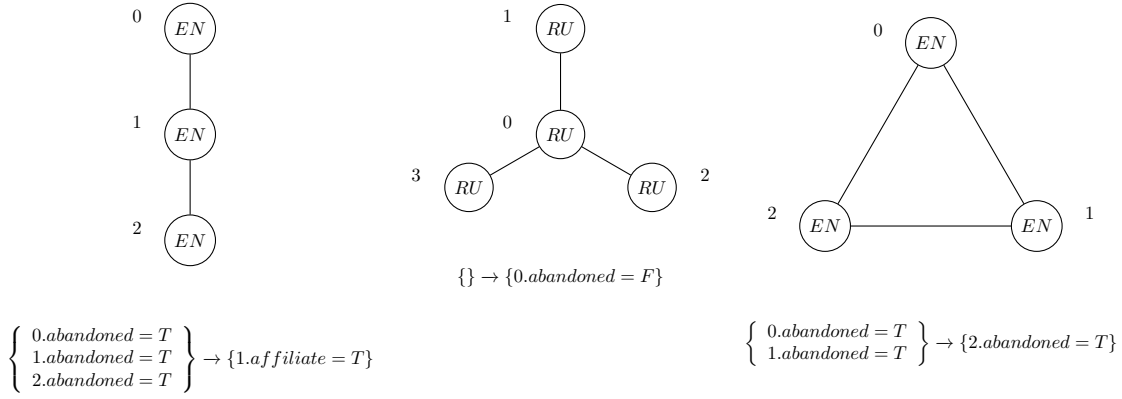
\begin{figure*}
\begin{center}
\begin{tikzpicture}[main/.style = {draw, circle}, scale=0.7, transform shape] 
\node[main] (p0) at (-7,2) {$EN$};
\coordinate [label=$0$] (pi0) at (-8,2);
\node[main] (p1) at (-7,0) {$EN$};
\coordinate [label=$1$] (pi1) at (-8,0);
\node[main] (p2) at (-7,-2) {$EN$};
\coordinate [label=$2$] (pi2) at (-8,-2);

\draw (p0) -- (p1) -- (p2);

\coordinate [label={$\left\{
\begin{array}{l l}
0.abandoned=T\\
1.abandoned=T\\
2.abandoned=T
\end{array}
\right\}
 \xrightarrow{} \{1.affiliate=T\}$}] (dependency_path) at (-7,-5);

\node[main] (s0) at (0,0) {$RU$};
\coordinate [label=$0$] (si0) at (-1,0);
\node[main] (s1) at (0,2) {$RU$};
\coordinate [label=$1$] (si1) at (-1,2);
\node[main] (s2) at (1.73,-1) {$RU$};
\coordinate [label=$2$] (si2) at (2.73,-1);
\node[main] (s3) at (-1.73,-1) {$RU$};
\coordinate [label=$3$] (si3) at (-2.73,-1);

\draw (s0) -- (s1);
\draw (s0) -- (s2);
\draw (s0) -- (s3);
\coordinate [label={$\{\} \xrightarrow{} \{0.abandoned=F\}$}] (dependency_star) at (0,-3);

\node[main] (t0) at (7,1.73) {$EN$};
\coordinate [label=$0$] (ti0) at (6,1.73);
\node[main] (t1) at (9,-1.73) {$EN$};
\coordinate [label=$1$] (ti1) at (10,-1.73);
\node[main] (t2) at (5,-1.73) {$EN$};
\coordinate [label=$2$] (ti2) at (4,-1.73);

\draw (t0) -- (t1) -- (t2);
\draw (t0) -- (t2);
\coordinate [label={$\left\{
\begin{array}{l l}
0.abandoned=T\\
1.abandoned=T
\end{array}
\right\} \xrightarrow{} \{2.abandoned=T\}$}] (dependency_triangle) at (7,-4.5);

\end{tikzpicture} 
\end{center}
\caption{GFDs-queries}
\label{queries}
\end{figure*}

\subsection{FastGFDs}

To improve subgraph matching performance, we propose an algorithm that uses ideas from~\cite{bi2016efficient}. This algorithm is identical to the baseline, except for Step 2. The improved Step 2 (graph pattern matching) is as follows:

\begin{enumerate}
    \item Building the \emph{compact path-index} (CPI) data structure:
    \begin{enumerate}
        \item First, we select vertex $v_q$ in query graph $Q$ and find a set of candidate vertices $v_{c_i}$ in data graph $G$ by searching vertices with the same label as $v_q$;
        \item then we filter out candidates that definitely do not match, by removing candidates with lower degree than $v_q$;
        \item then we enumerate vertices in a data graph with BFS, building candidate subgraphs; ``dead-end'' vertices (vertices that do not have the edges needed to continue the match) are removed.
        \item Since the CPI is built only for trees,
        we build a spanning tree for $Q$ starting from $v_q$, and then, if a match of this spanning tree is found, we additionally check remaining edges.
    \end{enumerate}
    \item Matching the graph pattern:
    \begin{enumerate}
        \item we search a minimal subgraph of $Q$ that contains a cycle~--- the \emph{core} of a query, $Q_c$;
        \item then we search for matches of $Q_c$ in $G$;
        \item then we extend found matches by enumerating remaining vertices in $Q$ and finding corresponding vertices in $G$.
    \end{enumerate}
\end{enumerate}

\section{Evaluation}\label{sec:evaluation}
 
In order to evaluate the proposed approach, we varied the embeddings enumeration subroutine. Overall, we have compared three implementations:
\begin{enumerate}
    \item Naive: a very naive GFD validation algorithm that uses the standard Boost method \verb+vf2_subgraph_iso+ for $Q$ embeddings enumeration. This implementation does not use the parallel scheme of the baseline method described in Section~\ref{sec:baseline-algo} in the first item of the list. This version is provided for completeness.
    
    \item Baseline: the original GFD validation algorithm with the parallel scheme, fully described in Section~\ref{sec:baseline-algo}. Since the authors did not specify the exact subgraph matching algorithm in the step 2, we employed the same \verb+vf2_subgraph_iso+ routine. To implement parallel processing, we used \verb+std::thread+, employing four of them during our experiments.

    \item Proposed (FastGFDs): the GFD validation algorithm that uses Core-First Decomposition and the Compact Path Index from paper~\cite{bi2016efficient} for subgraph enumeration. It does not rely on the parallel scheme, and instead processes the unpartitioned graph.
\end{enumerate}

We used the Twitch Gamers Social Network dataset~\cite{DBLP:journals/corr/abs-2101-03091}, available at the following link \url{http://snap.stanford.edu/data/twitch_gamers.html}. Since this dataset was too large for the employed hardware, we used only the first 20K nodes and all induced edges (92798).

This dataset describes the social relationships of Twitch users. Within the dataset, nodes correspond to accounts and edges denote followship relations. Accounts can have various properties, such as language (EN/RU), abandoned/not abandoned, etc. The affiliate property means that this account is controlled by another account.

For benchmarking purposes, we have implemented the following queries (see Fig.~\ref{queries}):
\begin{enumerate}
    \item Simple path: if there are three accounts, where one account ($1$) knows the other two, and all of them are abandoned, then $1$ is an affiliated account. At the same time, $0$ and $2$ accounts may or may not be friends.
    \item Star: if there are at least three friends, then the central account is not abandoned.
    \item Triangle: if an account $2$ has two friends, which are in turn friends with each other, and they have abandoned their accounts, then $2$ is abandoned too.
\end{enumerate}

Interestingly, all three dependencies hold, including the simple path one.

The hardware and software configuration was as follows: SAMSUNG NP350E5C-S07RU, Intel(R) Core(TM) i5-3210M (2 cores, 4 threads), 6GB RAM, x86\_64, 20.04.1-Ubuntu, g++ 9.4.0, boost 1.72.0.

These queries were selected since they are common in this graph, they are relatively small, and they result in a large number of embeddings. More complex queries almost immediately produce tiny or empty results.

The resulting run times are presented in Table~\ref{tab:results-time}. One can see that the naive approach loses to all other implementations for all three queries. The baseline approach is superior to it by 15\%--3x, depending on query. Thus, the parallel scheme proposed by Fan et al.~\cite{fan2016functional} provides positive speedup, even on modest laptop with low number of cores.

Next, the proposed approach consistently beats the naive up to 4 times (3.78x on average). It is also superior to the baseline, up to 3 times (2.6x on average). Note that the proposed approach does not rely on the parallel scheme, thus the sequential algorithm has beaten the parallel one. However, our experimental bench features a low number of cores, while in the original paper hundreds were used. 

Finally, it is interesting to note that the star query was the fastest out of all, the path was the slowest, and the triangle was in-between. At the same time, the improvement compared to the baseline varied greatly, and here the order is different: star (3.31x), path (3.06x), and triangle (1.45x).

Note that run times of the proposed approach include the time it takes to construct all the necessary CPI data structures. Therefore, in scenarios when the CPI can be precomputed in advance, the efficient run time can be even lower. However, the CPI is built for a particular query, and thus it cannot be easily reused.

Now, let us examine the memory consumption of the evaluated approaches, as shown in Table~\ref{tab:results-memory}. To do this, we measured the peak memory consumption of the whole application, using the \verb+VmPeak+ value of \verb+/proc+. It is evident that the naive implementation requires the least amount of memory to run. The baseline approach requires approximately 5 times more. This memory is used to store messages for the parallel scheme. Discarding the parallel scheme and employing sequential subgraph matching from paper~\cite{bi2016efficient} reduced it to the naive algorithm's level. Note that unlike run times, memory consumption does not depend on the studied pattern.

Based on these results, we can conclude the following:
\begin{enumerate}
    \item The parallel scheme yields positive results compared to the naive approach (with an average improvement of 1.74x), even on low-end hardware. However, this comes at a cost, as memory consumption increases five times.
    \item The modern sequential algorithm is significantly better than the parallel scheme (2.6x on average) with the \verb+vf2_subgraph_iso+ routine. Meanwhile, both its indexing time and the required space are negligible.
\end{enumerate}

\begin{table*}
\centering
\raisebox{2.2mm}[0pt][0pt]{%
\begin{minipage}{.49\textwidth}
	\centering
    \begin{tabular}{|c|c|c|c|}
        \hline
         & \multicolumn{3}{c|}{Query} \\
        \cline{2-4}
        \raisebox{1.5ex}[0cm][0cm]{Algorithm} & Star & Triangle & Path \\
        \hline
        Naive & 84 ms & 1689 ms & 5235 ms \\
        \hline
        Baseline & 73 ms & 685 ms & 4039 ms \\
        \hline
        FastGFDs & 22 ms & 470 ms & 1321 ms \\
        \hline
    \end{tabular}
    \caption{Run times of the considered implementations.}
    \label{tab:results-time}
\end{minipage}%
}
\hfill
\begin{minipage}{.49\textwidth}
	\centering
    \begin{tabular}{|c|c|c|c|}
        \hline
         & \multicolumn{3}{c|}{Query} \\
        \cline{2-4}
        \raisebox{1.5ex}[0cm][0cm]{Algorithm} & Star & Triangle & Path \\
        \hline
        Naive & 35956 kB & 35952 kB & 35952 kB \\
        \hline
        Baseline & 172656 kB & 172656 kB & 172652 kB \\
        \hline
        FastGFDs & 35956 kB & 35956 kB & 35952 kB \\
        \hline
    \end{tabular}
    \caption{Memory consumption of the considered implementations.}
    \label{tab:results-memory}
\end{minipage}
\end{table*}

\section{Threats to Validity}
\label{sec:threats}

Our work has the following threats to validity:

\begin{itemize}
    \item First of all, the CPU of our test bench had only two cores and supported only four threads. Therefore, the original algorithm is at a serious disadvantage, since it relies on parallel processing. Even slightly increasing the number of available cores could enable the original algorithm to reach its full potential and significantly improve its performance. Exploring this possibility is one of the directions of our future work.
    \item Another concern is the fact that the proposed algorithms were evaluated in an in-memory environment. If any of these algorithms run out of memory, disk swapping might occur, leading to a drastic decline in performance. In this regard, the original algorithm is in an inferior position, as it reaches the memory limit more quickly, given that it requires five times more memory than the others. Additionally, in our experiments, we used a small graph, resulting in a negligible CPI memory footprint. However, increasing graph size could change this outcome and potentially turn it into a serious limitation. To address this issue, further experiments with larger graphs are required. We also plan to explore this in our future work.
    \item Next, the performance of algorithms heavily depends on the query and dataset properties. In order to negate this threat, we have tried to evaluate these algorithms using several different queries.
    \item Finally, in our implementation, we relied on the Boost graph representation. Switching to another library for representation, or implementing a custom one, could affect the result. However, we believe that it might affect only the obtained ratios, but not the relative order of implementations. Exploration of this issue is also a matter of future work.
\end{itemize}

\section{Conclusion and Future Work}
\label{sec:conclusion}

In this paper, we described our open-source implementation of the graph functional dependency validation algorithm from study~\cite{fan2016functional} and an improved algorithm with fast graph pattern matching that targets specifically a low-end multi-threaded environment. Performance evaluation run on a real-life graph demonstrated that the improved algorithm shows up to three times performance (2.6x on average) increase over the
state-of-the-art algorithms. Employing the new subgraph matching algorithm also led to a fivefold reduction in memory consumption.

The source code of our C++ implementation and evaluation datasets can be found in the GitHub repository (\url{https://github.com/Mstrutov/Desbordante/pull/154}) of the Desbordante project~\cite{desbordanteRepo}.

There are several directions for future work:
\begin{itemize}
    \item Integrate the CPI-based subgraph matching into the parallel scheme of the original algorithm. However, the parallel scheme is designed to work with small graphs, while CPI is more suited for large graphs. Therefore, there should be a some kind of compromise on the graph size.
    \item Develop a special parallel scheme for the CPI-based subgraph matching. Since the previous approach might not yield positive results, a straightforward idea is to parallelize the CPI-based querying.
\end{itemize}

\section*{Acknowledgments}

We would like to thank Anna Smirnova for her help with the preparation of this paper.

\bibliographystyle{IEEEtran}
\bibliography{bibliography}

\begin{thebibliography}{10}
\providecommand{\url}[1]{#1}
\csname url@samestyle\endcsname
\providecommand{\newblock}{\relax}
\providecommand{\bibinfo}[2]{#2}
\providecommand{\BIBentrySTDinterwordspacing}{\spaceskip=0pt\relax}
\providecommand{\BIBentryALTinterwordstretchfactor}{4}
\providecommand{\BIBentryALTinterwordspacing}{\spaceskip=\fontdimen2\font plus
\BIBentryALTinterwordstretchfactor\fontdimen3\font minus
  \fontdimen4\font\relax}
\providecommand{\BIBforeignlanguage}[2]{{%
\expandafter\ifx\csname l@#1\endcsname\relax
\typeout{** WARNING: IEEEtran.bst: No hyphenation pattern has been}%
\typeout{** loaded for the language `#1'. Using the pattern for}%
\typeout{** the default language instead.}%
\else
\language=\csname l@#1\endcsname
\fi
#2}}
\providecommand{\BIBdecl}{\relax}
\BIBdecl

\bibitem{fan2016functional}
W.~Fan, Y.~Wu, and J.~Xu, ``Functional dependencies for graphs,'' in
  \emph{Proceedings of the 2016 international conference on management of
  data}, 2016, pp. 1843--1857.

\bibitem{10.1145/2882903.2915203}
\BIBentryALTinterwordspacing
T.~Papenbrock and F.~Naumann, ``A hybrid approach to functional dependency
  discovery,'' in \emph{Proceedings of the 2016 International Conference on
  Management of Data}, ser. SIGMOD '16.\hskip 1em plus 0.5em minus 0.4em\relax
  New York, NY, USA: Association for Computing Machinery, 2016, p. 821–833.
  [Online]. Available: \url{https://doi.org/10.1145/2882903.2915203}
\BIBentrySTDinterwordspacing

\bibitem{10.1145/3397198}
\BIBentryALTinterwordspacing
W.~Fan, C.~Hu, X.~Liu, and P.~Lu, ``Discovering graph functional
  dependencies,'' \emph{ACM Trans. Database Syst.}, vol.~45, no.~3, sep 2020.
  [Online]. Available: \url{https://doi.org/10.1145/3397198}
\BIBentrySTDinterwordspacing

\bibitem{mahdisoltani2014yago3}
F.~Mahdisoltani, J.~Biega, and F.~Suchanek, ``Yago3: A knowledge base from
  multilingual wikipedias,'' in \emph{7th biennial conference on innovative
  data systems research}.\hskip 1em plus 0.5em minus 0.4em\relax CIDR
  Conference, 2014.

\bibitem{bi2016efficient}
F.~Bi, L.~Chang, X.~Lin, L.~Qin, and W.~Zhang, ``Efficient subgraph matching by
  postponing cartesian products,'' in \emph{Proceedings of the 2016
  International Conference on Management of Data}, 2016, pp. 1199--1214.

\bibitem{yang2011br}
J.~Yang and W.~Jin, ``Br-index: An indexing structure for subgraph matching in
  very large dynamic graphs,'' in \emph{Scientific and Statistical Database
  Management: 23rd International Conference, SSDBM 2011, Portland, OR, USA,
  July 20-22, 2011. Proceedings 23}.\hskip 1em plus 0.5em minus 0.4em\relax
  Springer, 2011, pp. 322--331.

\bibitem{Desbordante}
\BIBentryALTinterwordspacing
G.~Chernishev, M.~Polyntsov, A.~Chizhov, K.~Stupakov, I.~Shchuckin, A.~Smirnov,
  M.~Strutovsky, A.~Shlyonskikh, M.~Firsov, S.~Manannikov, N.~Bobrov,
  D.~Goncharov, I.~Barutkin, V.~Shalnev, K.~Muraviev, A.~Rakhmukova,
  D.~Shcheka, A.~Chernikov, D.~Mandelshtam, M.~Vyrodov, A.~Saliou, E.~Gaisin,
  and K.~Smirnov, ``Desbordante: from benchmarking suite to high-performance
  science-intensive data profiler (preprint),'' 2023. [Online]. Available:
  \url{https://arxiv.org/abs/2301.05965}
\BIBentrySTDinterwordspacing

\bibitem{siek2001boost}
J.~G. Siek, L.-Q. Lee, and A.~Lumsdaine, \emph{The Boost Graph Library: User
  Guide and Reference Manual, The}.\hskip 1em plus 0.5em minus 0.4em\relax
  Pearson Education, 2001.

\bibitem{DBLP:journals/corr/abs-2101-03091}
\BIBentryALTinterwordspacing
B.~Rozemberczki and R.~Sarkar, ``Twitch gamers: a dataset for evaluating
  proximity preserving and structural role-based node embeddings,''
  \emph{CoRR}, vol. abs/2101.03091, 2021. [Online]. Available:
  \url{https://arxiv.org/abs/2101.03091}
\BIBentrySTDinterwordspacing

\bibitem{desbordanteRepo}
A.~Chernikov, ``Desbordante gfd pull request,''
  \url{https://github.com/Mstrutov/Desbordante/pull/154}, 2023.

\end{thebibliography}

\end{document}